\documentclass{article}
\usepackage[affil-it]{authblk}
\usepackage{amsmath, amsfonts, amsthm, amssymb}
\usepackage{commath}
\usepackage[usenames,dvipsnames]{color}
\usepackage{comment}
\usepackage{ifpdf}
\usepackage{setspace}
\usepackage[utf8]{inputenc}
\usepackage[english]{babel}
\usepackage{graphicx}  
\usepackage{breqn}
\usepackage{mathrsfs}
\usepackage{gensymb}
\usepackage{caption}
\usepackage{pgfplots}
\usepackage{cite}
\usepackage[left=.8in,right=.8in,top=.8in,bottom=.8in]{geometry}  
\numberwithin{equation}{section}
\newcommand*\colvec[3][]{
    \begin{pmatrix}\ifx\relax#1\relax\else#1\\\fi#2\\#3\end{pmatrix}
}

\begin{document}
\title{Where are the BTZ Black Hole Degrees of Freedom? \\
The Rotating Case}   
\author{Joseph M. Mitchell}
\affil{Department of Physics, University of California, Davis}

\date{\today}      
\maketitle

\begin{abstract}
\noindent Recent work has shown that the entropy of the non-rotating BTZ black hole can be derived from a dual conformal description at any spatial location. In this followup it is shown that a dual conformal description exists at any spatial location for the rotating BTZ black hole as well. As in the non-rotating case, two copies of the central charge $c^{\pm} = 3\ell/2G$ are recovered and the microcanonical Cardy formula yields the correct Bekenstein-Hawking entropy. 
\end{abstract}

\pagebreak 

\section{Introduction}
More than 40 years have passed since Hawking \cite{Hawk} and Bekenstein \cite{Beken} first showed that black holes are thermodynamic objects with temperatures and entropies. However, the explicit details of the states accounting for these thermodynamic properties has remained a mystery. Perhaps most puzzling is the existence of numerous microscopic descriptions that all give the correct entropy despite counting physically different states. Additionally, Hawking's original derivation does not rely on any candidate theory of quantum gravity at all. One possible explanation for this universality is that the degrees of freedom might be governed by a classical symmetry that is independent of the details of the quantum theory. As first suggested in \cite{ConformOrigin}, a two dimensional dual conformal description is a plausible candidate for this approach. 

This idea was fully implemented by Strominger \cite{Strominger} (and independently by Birmingham et al. \cite{Birmingham_CFT}) for the BTZ black hole \cite{BTZsolution} by imposing Brown-Henneaux boundary conditions \cite{BrownADS}. However, this places the dual CFT at spatial infinity. A physically more appealing location for the degrees of freedom is the event horizon, especially for higher dimensional black holes \cite{SolodukhinBCHorizon, CarlipBCHorizon, CarlipBCHorizon2, CarlipReply, Park_Comment, Park_Canonical, Park_NHCFT}. This still seems inadequate for 2+1 dimensions because the classical theory contains no local degrees of freedom. However, recent work has shown that a dual conformal description for the non-rotating BTZ black hole exists at any spatial location \cite{NonrotatingCase}. A similar result was obtained by Comp\`ere et al. in the asymptotically de Sitter context, although with some key differences in the calculation \cite{DeSitterComp}. \footnote{In particular, the derivations make use of different coordinate systems and the boundary conditions appear to be different. Additionally, in \cite{DeSitterComp} the entropy is not computed.}

There is also ambiguity in what boundary conditions to impose. Although most approaches fix the intrinsic geometry of the boundary, there seems to be no reason one could not impose conditions on the extrinsic curvature. For example, by working in the Chern-Simons formulation, Carlip has shown that different choices of boundary conditions can still lead to the correct entropy of the BTZ black hole\cite{Whatwedontknow}. One might be tempted to conclude that, at least in the Chern-Simons formulation, the choice of boundary conditions does not affect the state counting. However, as noted by Carlip, the derivations in \cite{Whatwedontknow} are sensitive to the location of the boundary. Since the results of \cite{NonrotatingCase} hold at any spatial location, there is hope that it could shed some light on this issue. 

The main purpose of this work is to extend the analysis of \cite{NonrotatingCase} to the full rotating BTZ black hole. In particular, the central charges $c^{\pm} = 3\ell/2G$ of Brown and Henneaux, as well as the conformal weights obtained by Strominger, are recovered. The microcanonical Cardy formula is then used to compute the correct Bekenstein-Hawking entropy. 


\section{A Dual CFT at Any Location}
Let us begin with the full rotating BTZ black hole \cite{BTZsolution} written in the familiar Schwarzschild-like coordinates

\begin{equation}
	ds^2 = - (r^2 - r_+^2 - r_-^2) dt^2 -2r_+r_- dtd\bar{\phi} +  \frac{ \ell^2 r^2 dr^2}{(r^2 - r_+^2)(r^2 - r_-^2)} + r^2 d\bar{\phi}^2
\end{equation}	
where the cosmological constant $\Lambda = - 1/\ell^2$ and $r_{\pm}$ are the inner and outer horizons:

\begin{equation}
	r_{\pm}^2 = 4GM\ell^2 \Bigg(1 \pm \Bigg[1 - \Bigg(\frac{J}{M\ell}\Bigg)\Bigg]^{1/2}\Bigg)
\end{equation}	
Here $M$ and $J$ correspond to the mass and angular momentum of the black hole, respectively:
\begin{equation}
	M = \frac{r_+^2 + r_-^2}{8G\ell^2}, \ \ \ \ J = \frac{r_+r_-}{4G\ell}, \ \ \ \ \text{with} \  (\abs{J} \leq M\ell)
\end{equation}

As in \cite{NonrotatingCase} for the non-rotating case, it is useful to put the metric into a form adaptable to circular null coordinates. This was originally motivated by noting that the asymptotic symmetries of Brown and Henneaux depend on the combinations $t \pm \bar{\phi}$ \cite{BrownADS}, which are null at spatial infinity. First, I change to an angular coordinate that co-rotates with the black hole horizon:

\begin{equation}
		\phi = \bar{\phi} - \frac{r_-}{r_+} t
\end{equation}
so

\begin{equation}
	ds^2 = - \frac{r_+^2 - r_-^2}{r_+^2}(r^2 - r_+^2) dt^2 + 2\frac{r_-}{r_+}(r^2 - r_+^2)dtd\phi +  \frac{ \ell^2 r^2 dr^2}{(r^2 - r_+^2)(r^2 - r_-^2)} + r^2 d\phi^2
\end{equation}
Periodicity in $\bar{\phi}$ implies the identification $\phi \sim \phi + 2\pi$. I make one more change of coordinates:
\begin{equation}
	\begin{split}
		r^2& =  (r_+^2 - r_-^2)e^{2\rho/\ell - 2r_+ \tau/\ell - 2r_-\phi/\ell} + r_-^2 \\
		t &= \frac{r_+^2}{r_+^2 - r_-^2} \Big(\tau + 2\frac{r_-}{r_+}\phi - \frac{\ell}{2r_+} \ln(1 - e^{-2\rho/\ell + 2r_+ \tau/\ell + 2r_-\phi/\ell})\Big) \\
	\end{split}
\end{equation}
This puts the metric into a fairly simple form, in which the surfaces of constant $\rho$ are conformally flat:
\begin{equation}
	ds^2 = e^{2\rho/\ell - 2r_+ \tau/\ell - 2r_-\phi/\ell}\Big(-r_+^2 d\tau^2 - 2r_+r_- d\tau d\phi + (r_+^2 - r_-^2)d\phi^2\Big) + d\rho^2
\end{equation}	
Here I introduce the notation
\begin{equation}
	\tau^{\pm} = \frac{r_+}{r_+ \pm r_-} \tau
\end{equation}	
One can easily check that the combinations $\tau^{\pm}  \pm \phi$ are null on any surface of constant $\rho$. Also note that when $J \rightarrow 0$ (equivalently, $r_- \rightarrow 0$), the metric (2.7) and the null coordinates $\tau^{\pm} \pm \phi$ reduce to those obtained in \cite{NonrotatingCase}. 

I now impose the following conditions on the set of allowed diffeomorphisms:
\begin{equation}
	\begin{split}
		\mathcal{L}_{\xi} g_{\tau \tau} &= 0  \ \Rightarrow \ \xi^{\rho} = r_+ \xi^{\tau} + r_- \xi^{\phi} - \ell \partial_{\tau}\xi^{\tau} - \frac{r_-}{r_+} \partial_{\tau} \xi^{\phi} \\
		\mathcal{L}_{\xi} g_{\phi \phi} &= 0  \ \Rightarrow \ \xi^{\rho} = r_+ \xi^{\tau} + r_- \xi^{\phi} - \ell \partial_{\phi}\xi^{\phi} + \frac{r_+ r_-}{r_+^2 - r_-^2} \partial_{\phi} \xi^{\tau}\\	
		\mathcal{L}_{\xi} g_{\phi \tau} &= 0 \  \Rightarrow \ \xi^{\rho} = r_+ \xi^{\tau} + r_- \xi^{\phi} -\frac{\ell}{2} \partial_{\phi}\xi^{\phi} - \frac{\ell}{2} \partial_{\tau}\xi^{\tau} - \frac{r_+}{2r_-}\partial_{\phi} \xi^{\tau} + \frac{r_+^2 - r_-^2}{2r_+ r_-} \partial_{\tau} \xi^{\phi} \\	
		\mathcal{L}_{\xi} g_{\rho \rho} &= 0 \ \Rightarrow \ \partial_{\rho} \xi^{\rho} = 0 
	\end{split}		
\end{equation}	
while allowing $\mathcal{O}(1)$ changes in $g_{\rho \tau}$ and $g_{\rho \phi}$. Diffeomorphisms preserving these boundary conditions take the form
\begin{equation}
	\begin{split}
		\xi^{\pm \tau} &= \frac{r_+^2 - r_-^2}{2r_+^2} T^{\pm}(\tau^{\pm} \pm \phi)\\
		\xi^{\pm \rho} &= \frac{r_+ \pm r_-}{2} T^{\pm}(\tau^{\pm} \pm \phi) \mp \frac{\ell}{2}\partial_{\phi} T^{\pm}(\tau^{\pm} \pm \phi) \\
		\xi^{\pm \phi} &= \pm \frac{r_+ \pm r_-}{2 r_+} T^{\pm}(\tau^{\pm} \pm \phi) 
	\end{split}	
\end{equation}
where $T^{\pm}$ depends only on the combinations $\tau^{\pm} \pm \phi$ and is otherwise arbitrary. The vector field has been normalized such that its commutator is
\begin{equation}
	\begin{split}
		[\xi^{\pm}(T^{\pm}_1), \xi^{\pm}(T^{\pm}_2)] &= \xi^{\pm}(T^{\pm}_1\partial_{\phi}T^{\pm}_2 - T^{\pm}_2\partial_{\phi}T^{\pm}_1)\\
		[\xi^{+}(T^{+}_1), \xi^{-}(T^{-}_2)] &= 0
	\end{split}
\end{equation}		
In terms of modes $T_n^{\pm} = e^{in(\tau^{\pm} \pm \phi)}$ this becomes
\begin{equation}
	\begin{split}
		[\xi^{\pm}_n, \xi^{\pm}_m] &= i(m-n)\xi^{\pm}_{n + m}\\
		[\xi^{+}_n, \xi^{-}_m] &= 0
	\end{split}
\end{equation}		
So the commutator forms a pair of commuting Witt algebras. As expected, when $J \rightarrow 0$ the vector field (2.10) reduces to what was obtained in \cite{NonrotatingCase}.

I will now switch to the canonical formalism \cite{ADM}, reviewed briefly in the appendix, in order to extract any nonzero central charges appearing in the algebra of the generators. From the general ADM form of the metric

\begin{equation}
	ds^2 = -N^2 dt^2 + q_{ij}(dx^i + N^i dt)(dx^j + N^j dt)
\end{equation}
it is easy to read off the lapse $N$ and shift $N^i$ from (2.7):

\begin{equation}
	N = \frac{r_+^2}{\sqrt{r_+^2 - r_-^2}}e^{\rho/\ell - r_+\tau/\ell - r_-/\ell \phi}, \ \ \ N^{\phi} = -\frac{r_+ r_-}{r_+^2 - r_-^2}
\end{equation}
The only non-vanishing component of the canonical momentum (A.2) is

\begin{equation}
	\pi^{\rho \rho} = \frac{r_+}{\ell}
\end{equation}		
Spacetime diffeomorphisms $\xi^{\mu}$ are replaced by the surface deformation parameters $(\xi^{\perp}, \hat{\xi}^i)$, related through \cite{TeitSDParameter}:

\begin{equation}
	\xi^{\perp} = N\xi^t, \ \ \hat{\xi}^i = \xi^i + N^i \xi^t
\end{equation}

The boundary conditions (2.9) are imposed on the full spacetime metric $g_{\mu \nu}$ under Lie transport. Instead, boundary conditions must be imposed on the canonical variables $(q_{ij}, \pi^{ij})$ under Hamiltonian transport (A.5):
\begin{equation}
	\begin{split}
		&\delta_{\xi} q_{\rho \rho} = 0 \ \Rightarrow \ \partial_{\rho} \hat{\xi}^{\rho} = 0 \\ 
		&\delta_{\xi} q_{\phi \phi} = 0 \ \Rightarrow \ \hat{\xi}^{\rho} = \frac{\ell \pi}{\sqrt{q}} \xi^{\perp} + r_- \hat{\xi}^{\phi} - \ell \partial_{\phi} \hat{\xi}^{\phi}
	\end{split}
\end{equation}	
where I have temporarily neglected boundary terms. The lapse $N$ and shift $N^i$ are not dynamical, so they still transform according to the Lie derivative (2.9):
\begin{equation}
	\begin{split}
		\delta_{\xi} N \ &= 0 \ \Rightarrow \ N \hat{\xi}^{\rho} = N^{\phi}\partial_{\phi} \xi^{\perp} + r_- N\hat{\xi}^{\phi} -\ell \partial_{\tau} \xi^{\perp} \\
		\delta_{\xi} N^{\phi} &= 0 \ \Rightarrow \partial_t \hat{\xi}^{\phi} - N^{\phi} \partial_{\phi} \hat{\xi}^{\phi} = Nq^{\phi \phi} \partial_{\phi} \xi^{\perp} - q^{\phi \phi}\xi^{\perp}\partial_{\phi} N  \\
	\end{split}
\end{equation}	
Changes in $N^{\rho}$ and $q_{\rho \phi}$ are still $\mathcal{O}(1)$. As in the non-rotating case, since variations in $q_{\rho \phi}$ are not required to vanish, the absent cross-term of the ADM momentum can be held fixed:

\begin{equation}
	\delta_{\xi} \pi^{\rho \phi} = 0 \Rightarrow \pi^{\rho \rho} \partial_{\rho} \hat{\xi}^{\phi} - \sqrt{q} q^{\rho \rho} q^{\phi \phi}\Big(\partial_{\rho} \partial_{\phi} -\frac{1}{\ell} \partial_{\phi}\Big) \xi^{\perp} = 0
\end{equation}
Surface deformations preserving (2.17)-(2.19) take the form
\begin{equation}
	\begin{split}
		\xi^{\pm \perp} &= \frac{1}{2}\sqrt{r_+^2 - r_-^2}e^{\rho/\ell - r_+\tau/\ell - r_-\phi/\ell} \ T^{\pm}(\tau^{\pm} \pm \phi) \\
		\hat{\xi}^{\pm \rho} &= \frac{r_+ \pm r_-}{2} T^{\pm}(\tau^{\pm} \pm \phi) \mp \frac{\ell}{2} \partial_{\phi} T^{\pm}(\tau^{\pm} \pm \phi) \\
		\hat{\xi}^{\pm \phi} &= \pm \frac{1}{2} T^{\pm}(\tau^{\pm} \pm \phi) 
	\end{split}	
\end{equation}
From (2.14) and (2.16), the surface deformations (2.20) clearly agree with the diffeomorphisms (2.10). Under these transformations, changes in $\pi^{\rho \rho}$ are $\mathcal{O}(1)$ while variations in $\pi^{\phi \phi}$ vanish. However, since $q_{\phi \phi}$ is already being held fixed I do not impose that $\pi^{\phi \phi}$ be held fixed as a boundary condition. This also occurred in \cite{NonrotatingCase} for the non-rotating case, in which it was suggested that this may arise from some combination of the boundary conditions.

There is one final subtlety that must be take into account. Note that the trace of the ADM momentum $\pi$ and the determinant of the spatial metric $q$ appear in the second line of (2.17). One can see fom (2.18)-(2.19) that neither $\xi^{\perp}$ nor $\hat{\xi}^{\phi}$ depend on $\pi$ or $q$, so (2.17) implies that $\hat{\xi}^{\rho}$ depends on both quantities. As discussed in the appendix, when the surface deformations have dependence on the canonical variables, one cannot use their usual Lie bracket \cite{TeitSDParameter}. One must instead use the full surface deformation bracket (A.9), which, after a somewhat tedious calculation, yields a pair of commuting Witt algebras:

\begin{equation}
	\begin{split}
		&\{\xi^{\pm}_n,\xi^{\pm}_m\}^{\perp} = i(m - n)\xi^{\pm \perp}_{n+m} \ \ \ \ \{\xi^{+}_n,\xi^{-}_m\}^{\perp} = 0 \\
		&\{\xi^{\pm}_n,\xi^{\pm}_m\}^i \ = i(m - n)\hat{\xi}^{\pm i}_{n+m} \ \ \ \  \{\xi^{+}_n,\xi^{-}_m\}^i \ = 0 \\
	\end{split}
\end{equation}		
where the mode expansion of $T^{\pm}$ has been used for convenience. At this point one might worry that the surface deformations (2.20) could have additional dependence on the canonical variables. Although this is indeed the case, the algebra (2.21) is unaffected. From (2.20), clearly $\xi^{\perp}$ depends on $\rho$, which, from the form of the metric (2.7), is the proper distance on a constant time slice, and is therefore metric dependent. However, it is easy to see from (A.9) that this adds to the Lie bracket terms proportional to $\{q_{\rho \rho}, H[\xi]\}$, which vanish due to the choice of boundary conditions. The same conclusion can be reached for any possible dependence on $q_{\phi \phi}$ as well.

Finally, I now turn to the general form for the central terms in the canonical formalism (A.12) as derived by Carlip in \cite{Whatwedontknow}. The relevant three-derivative terms are

\begin{equation}
	K[\xi, \eta] = ... - \frac{1}{8\pi G} \int{ d\phi \sqrt{\sigma} n^k (D_m \hat{\xi}_k D^m \eta^{\perp} - D_m \hat{\eta}_k D^m \xi^{\perp})}
\end{equation}	
where $n^k$ is the unit normal, $D_m$ is the spatial covariant derivative compatible with the spatial metric $q_{ij}$ and $\sigma$ is the determinant of the induced metric on the boundary $\sigma_{ij}$. Evaluating (2.22) with the surface deformations (2.20) yields

\begin{equation}
	K[\xi^{\pm}(T_1^{\pm}), \xi^{\pm}(T_2^{\pm})] = ... \mp \frac{\ell}{32 \pi G} \int{d\phi (\partial_{\phi} T_1^{\pm}\partial_{\phi}^2 T_2^{\pm} - \partial_{\phi} T_2^{\pm}\partial_{\phi}^2 T_1^{\pm})} \\
\end{equation}	
from which one can read off the central charges \cite{CFTyellowpages}:

\begin{equation}
	c^{\pm} = \frac{3\ell}{2G}
\end{equation}
These central charges are the same as obtained by Brown and Henneaux. As in \cite{NonrotatingCase}, however, the analysis here holds at any spatial location. 

\section{The Conformal Weight and Entropy}
In this section the entropy of the rotating BTZ black hole is computed from the microcanonical version of the Cardy formula \cite{Cardy1, Cardy2}:

\begin{equation}
	S = 2\pi \sqrt{\frac{c}{6}\Big(\Delta - \frac{c}{24}\Big)}
\end{equation}
where the conformal weight $\Delta$ is determined by the boundary term $B[\xi]$. As discussed in the Appendix, the boundary term must be chosen to cancel the boundary variation of the Hamiltonian \cite{HamBterms}. For the case at hand, the non-vanishing terms in this variation (A.10) are

\begin{equation}
	\delta H[\xi] = ... -\frac{1}{16\pi G} \int d\phi \Big\{ \sqrt{\sigma} \Big[\sigma^{\phi \phi} n^{\rho} \xi^{\perp} \Big(D_{\phi} \delta q_{\rho \phi} - D_{\rho} \delta q_{\phi \phi}\Big) - D_{\phi} \xi^{\perp} n^{\rho} \sigma^{\phi \phi} \delta q_{\rho \phi} \Big] + 2 \hat{\xi}^{\rho} \delta \pi^{\rho}_{\rho} \Big\}
\end{equation}	
where I have used the form of the metric (2.7) along with the boundary conditions (2.17)-(2.19). One can rewrite the first two terms as the variation of the mean extrinsic curvature of the boundary $k = \sigma^{\phi \phi} D_{\phi}n_{\phi}$ treated as a submanifold of the spatial slice. One can use $\delta (q_{i j} n^i n^j) = 0$ in order to rewrite the third term as the variation of the normal $\delta n^k = -\frac{1}{2} q^{k i} n^j \delta q_{i j}$. The last term requires some extra care because, as already mentioned, $\hat{\xi}^{\rho}$ is dependent on the canonical variables:

\begin{equation}
		\delta_{\xi} (\hat{\xi}^{\rho} \pi_{\rho}^{\rho}) = \hat{\xi}^{\rho} \delta\pi_{\rho}^{\rho} + \pi_{\rho}^{\rho} \frac{\ell}{\sqrt{q}} \xi^{\perp} \delta\pi
\end{equation}	
where the second term comes from the variation of $\hat{\xi}^{\rho}$ and use was made of (2.17). I have also used 
\begin{equation}
	\delta_{\xi} \sqrt{q} = \frac{1}{2\sqrt{q}} q^{ij}\delta q_{ij}
\end{equation}
which vanishes for the metric and boundary conditions considered here. So the boundary term is

\begin{equation}
	B[\xi] = \frac{1}{8\pi G} \int{d\phi\Big[\sqrt{\sigma} \Big(n^m \partial_m \xi^{\perp} - k\xi^{\perp} \Big) + \hat{\xi}^{\rho} \pi^{\rho}_{\rho} - \frac{\ell}{2\sqrt{q}} \pi^2 \xi^{\perp} \Big]} + B_0
\end{equation}
where $B_0$ is an arbitrary constant. For zero modes $T^{\pm}_0 = 1$ this reduces to
\begin{equation}
	B[T^{\pm}_0]  = \frac{(r_+ \pm r_-)^2}{16G\ell} + B_0
\end{equation}	
I now fix the arbitrary constant $B_0$ using the same method as Strominger \cite{Strominger}: when $r_+^2 \rightarrow -\ell^2$ and $r_- \rightarrow 0$ the BTZ metric (2.1) becomes that of AdS$_3$, so if one requires AdS$_3$ to have vanishing conformal weights the boundary term becomes
\begin{equation}
B[T^{\pm}_0] = \Delta^{\pm} = \frac{(r_+ \pm r_-)^2}{16G\ell} + \frac{\ell}{16G}
\end{equation}
These conformal weights and the central charges (2.24) can now be used with the microcanonical Cardy formula to yield the correct Bekenstein-Hawking entropy:
\begin{equation}
	S = 2\pi \sqrt{\frac{c^+}{6}\Big(\Delta^+ - \frac{c^+}{24}\Big)} +  2\pi \sqrt{\frac{c^-}{6}\Big(\Delta^- - \frac{c^-}{24}\Big)}= \frac{2\pi r_+}{4G}
\end{equation}
As in \cite{NonrotatingCase} for the non-rotating case, the calculation here holds at all spatial locations. 


\section{A Stationary Observer in the $(\tau, \rho, \phi)$ Coordinate System}
I will now explore some properties of the $(\tau, \rho, \phi)$ coordinate system since, as far as I am aware, the form of the metric (2.7) has not been used before. I begin by noting that the induced metric on surfaces of constant $\rho$ is conformally flat. If one switches to null coordinates
\begin{equation}
	u = \tau^+ + \phi, \ \ \ v = \tau^- - \phi
\end{equation}
the metric becomes
\begin{equation}
		ds^2 = -4\lambda^+ \lambda^-e^{(2\rho/\ell - 2\lambda^+ u/\ell - 2\lambda ^- v)} dudv + d\rho^2, \ \ \ \lambda^{\pm} = r_+ \pm r_-
\end{equation}
Any transformation of the form
\begin{equation}
	u \rightarrow f(u), \ \ \ v \rightarrow g(v)
\end{equation}
preserves the conformal structure. It is interesting that the dual CFT is fairly straightforward to obtain in a coordinate system with this structure. Also suggestive is that the boundary conditions (2.9) fix the induced metric on a surface of constant $\rho$. 

One might therefore wonder what $\rho$ represents physically. First, as already mentioned, $\rho$ is a proper distance coordinate. However, from (2.6), it is easy to see that an observer at constant $\rho$ does not remain at a fixed distance from the black hole. Remaining at a fixed location requires 
\begin{equation}
	\rho - r_+\tau - r_-\phi = \text{constant}
\end{equation}
In particular, the event horizon itself is located at
\begin{equation}
	\rho - r_+ \tau - r_- \phi = 0
\end{equation}
So the horizon is ``moving outward." This leads to the question of what an observer at constant $\rho$ experiences. The 4-velocity can easily be read off of the metric (2.7)
\begin{equation}
u^{\mu} = \Big(\frac{1}{r_+}e^{-\rho/\ell + r_+\tau/\ell + r_-\phi/\ell}, 0, 0\Big)
\end{equation}
and corresponds to a proper acceleration
\begin{equation}
	a^{\mu} = u^{\nu} \nabla_{\nu} u^{\mu} = (0, 1/\ell, 0) \ \Rightarrow \ a^2 = g_{\mu \nu} a^{\mu} a^{\nu} = \frac{1}{\ell^2}
\end{equation}	
So an observer at fixed $\rho$ and $\phi$ experiences a constant radial acceleration proportional to the cosmological constant. 

In order to see this motion graphically, refer back to the form of the metric (2.5), which co-rotates with the black hole horizon. It is useful to use a tortoise-like radial coordinate:
\begin{equation}
	r^2 = (r_+^2 - r_-^2)\coth^2(r_+r_{*} / \ell) + r_-^2	
\end{equation}
The metric then takes the form

\begin{equation}
	ds^2 = \frac{r^2_+}{\sinh^2(r_+r_* / \ell)} (-dt^2 + 2\frac{r_-}{r_+} dtd\phi+ dr^2_*) + ((r_+^2 - r_-^2) \coth^2(r_+r_* / \ell) + r_-^2)d\phi^2
\end{equation}
where I have also rescaled the time coordinate 
\begin{equation}
	t \rightarrow \frac{r_+^2}{r_+^2 - r_-^2} t
\end{equation}	
This choice of radial coordinate has the expected behavior: radial null geodesics are straight lines at $45 \degree$ angles, the black hole horizon is located at $r_{*} \rightarrow \infty$ and one approaches spatial infinity as $r_{*} \rightarrow 0$. From (2.6), one can plot the motion of an observer at constant $\rho$ in the $r_{*} - t$ plane. For simplicity, I have set $\phi = 0$.

\begin{center}
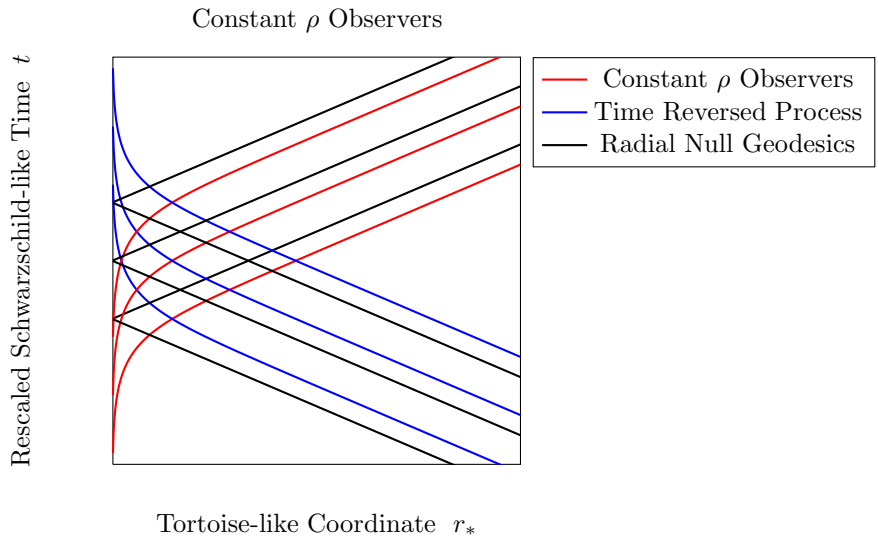

\begin{tikzpicture}
	\begin{axis}[
        xmin=0.01, xmax=6, 
        ticks=none,
       xticklabel style={/pgf/number format/.cd,fixed}, 
        width=7cm, height=7cm,
        ymin=-5,      
        ymax= 9,      
        axis background/.style={fill=white},
        ylabel={Rescaled Schwarzschild-like Time \ $t$},
        xlabel={Tortoise-like Coordinate \ $r_*$},
        title={Constant $\rho$ Observers},
        legend pos = outer north east
     ]
	\addplot[samples = 500, domain = 0.01:6, thick, red]{ln(sinh(x))};
	\addplot[samples = 500, domain = 0.01:6, thick, blue]{-ln(sinh(x))};
	\addplot[samples = 500, domain = 0.01:6, thick, black]{x};
	\addplot[samples = 500, domain = 0.01:6, thick, red]{ln(sinh(x)) + 2};
	\addplot[samples = 500, domain = 0.01:6, thick, blue]{-ln(sinh(x)) + 2};
	\addplot[samples = 500, domain = 0.01:6, thick, red]{ln(sinh(x)) + 4};
	\addplot[samples = 500, domain = 0.01:6, thick, blue]{-ln(sinh(x)) + 4};
	\addplot[samples = 500, domain = 0.01:6, thick, black]{-x};
	\addplot[samples = 500, domain = 0.01:6, thick, black]{x+2};
	\addplot[samples = 500, domain = 0.01:6, thick, black]{-x+2};
	\addplot[samples = 500, domain = 0.01:6, thick, black]{x+4};
	\addplot[samples = 500, domain = 0.01:6, thick, black]{-x+4};
	\legend{Constant $\rho$ Observers, Time Reversed Process, Radial Null Geodesics}
	\end{axis}
\end{tikzpicture}
\captionof{figure}{Red curves represent constant $\rho$ observers for various values of $\rho$. Blue curves are the time reversed process and black lines are ingoing and outgoing radial null geodesics.}\label{name}
\end{center}

So an observer at fixed $\rho$ corresponds to radial infall into the black hole. Considering this observer experiences a constant radial acceleration proportional to the cosmological constant, it seems likely that the form of the metric (2.7) is well-adapted to freely falling observers. 


\section{Conclusion}
The work here successfully extends the ideas in \cite{NonrotatingCase} to the full rotating BTZ black hole. In particular, a dual conformal description is obtained without taking either a near horizon or $r \rightarrow \infty$ limit. The central charges of Brown and Henneux, as well as the conformal weights first calculated by Strominger, are recovered. This suggests that in 2+1 dimensions the black hole degrees of freedom exist at any spatial location. This result is physically appealing  because 2+1 dimensional general relativity has no propagating degrees of freedom. However, some limitations remain. It would be ideal to have a physical explanation of the boundary conditions imposed here, as an important first step towards extending this work to higher dimensional black holes in the near horizon limit. It might also prove to be an important step towards developing a more general dual CFT procedure instead of studying one black hole solution at a time. Work in this direction will be explored in the future. 

\subsection*{Acknowledgements} 
I am very grateful to Steven Carlip for all of his helpful insights and suggestions throughout the course of this project. This work was supported in part by Department of Energy grant DE-FG02-91ER40674.


 \renewcommand{\theequation}{A.\arabic{equation}}
 \setcounter{equation}{0} 
 \section*{APPENDIX \ \ \ \ Relevant Aspects of Hamiltonian Gravity} 
An arbitrary metric in an n-dimensional spacetime can be written in ADM form as \cite{EffectiveConform}

\begin{equation}
	ds^2 = -N^2 dt^2 + q_{ij}(dx^i + N^i dt)(dx^j + N^j dt)
\end{equation}
where the lapse $N$ and shift $N^i$ are Lagrange multipliers. One treats as dynamical the spatial metric $q_{ij}$ and its canonically conjugate momentum $\pi^{ij}$. The conjugate momentum is related to the extrinsic curvature $K_{ij}$ through

\begin{equation}
	\pi^{ij} = \sqrt{q}(K^{ij} - q^{ij}K)
\end{equation}
where $K$ is the trace of $K_{ij}$ and I have set $16\pi G = 1$. Units will be restored when useful.

In the full spacetime picture a diffeomorphism is generated by a vector field $\xi^{\mu}$, but in this formulation symmetries are generated by the Hamiltonian:

\begin{equation}
	H[\xi] = \int{d^{(n-1)}x (\xi^{\perp} \mathcal{H} + \hat{\xi}^i \mathcal{H}_i)}
\end{equation}
The Hamiltonian constraint $\mathcal{H}$ and the momentum constraint $\mathcal{H}^i$ are

\begin{equation}
	\mathcal{H} = \frac{1}{\sqrt{q}}\Big(\pi^{ij}\pi_{ij} - \frac{1}{n - 2}\pi^2\Big) - \sqrt{q}\Big({}^{(n-1)}R - 2\Lambda\Big), \ \ \ \ \ \mathcal{H}^i = -2D_j \pi^{ij}
\end{equation}
where ${}^{(n - 1)}R$ is the spatial Ricci scalar and $D_j$ is the spatial covariant derivative. On a manifold without boundary, the transformation generated by the Hamiltonian is

\begin{equation}
	\begin{split}
		\{H[\xi], q_{ij}\} =& -\frac{2}{\sqrt{q}} \xi^{\perp} \Big(\pi_{ij} - \frac{1}{n - 2} q_{ij} \pi \Big) - \Big(D_i \hat{\xi}_j + D_j \hat{\xi}_i\Big) \\
		\{H[\xi], \pi^{ij}\} =& \ \sqrt{q} \ \xi^{\perp} \Big({}^{(n - 1)}R^{ij} - \frac{1}{2} q^{ij} \Big({}^{(n-1)}R - 2\Lambda\Big)\Big) + \frac{2}{\sqrt{q}} \xi^{\perp} \Big(\pi^{ik}\pi_k^j - \frac{1}{n-2} \pi\pi^{ij}\Big) \\
		&-\frac{1}{2\sqrt{q}} \xi^{\perp} q^{ij}\Big(\pi^{k\ell}\pi_{k\ell} - \frac{1}{n-2} \pi^2\Big) - \sqrt{q}\Big(D^iD^j\xi^{\perp} - q^{ij} D_kD^k\xi^{\perp}\Big)\\
		&-D_k\Big(\hat{\xi}^k\pi^{ij}\Big) + \pi^{ik}D_k\hat{\xi}^j + \pi^{ik}D_k\hat{\xi}^i\\
	\end{split}
\end{equation}	
On-shell the above transformation is equivalent to a spacetime diffeomorphism $\xi^{\mu}$ related to the surface deformation parameters $(\xi^{\perp}, \hat{\xi}^i)$ by \cite{TeitSDParameter}

\begin{equation} 
	\xi^{\perp} = N\xi^t, \ \ \ \ \hat{\xi}^i = \xi^i + N^i \xi^t
\end{equation}
	
One can also show that the algebra of the generators closes,	

\begin{equation}
	\{H[\xi], H[\eta]\} = H[\{\xi,\eta\}_{SD}]
\end{equation}
with the Lie bracket of surface deformations given by \cite{TeitSDParameter}

\begin{equation}
	\begin{split} 
		\{\xi, \eta\}^{\perp}_{SD} &= \hat{\xi}^i D_i \eta^{\perp} - \hat{\eta}^i D_i \xi^{\perp} \\
		\{\xi,\eta\}^i_{SD} &= \hat{\xi}^k D_k \eta^i - \hat{\eta}^k D_k \hat{\xi}^i + q^{i k} (\xi^{\perp} D_k \eta^{\perp} - \eta^{\perp} D_k \xi^{\perp}) \\
	\end{split}	
\end{equation}
The bracket (A.8) assumes the surface deformation parameters have no dependence on the canonical variables. When this assumption fails, one must use the full Lie bracket \cite{BrownADS, TeitSDParameter}:

\begin{equation}
	\begin{split} 
		\{\xi, \eta\}_{full}^{\perp} &= \hat{\xi}^i D_i \eta^{\perp} - \hat{\eta}^i D_i \xi^{\perp} + \{H[\xi], \eta^{\perp}\}_{PB} - \{H[\eta], \xi^{\perp}\}_{PB}\\
		\{\xi,\eta\}_{full}^i &= \hat{\xi}^k D_k \eta^i - \hat{\eta}^k D_k \hat{\xi}^i + q^{i k} (\xi^{\perp} D_k \eta^{\perp} - \eta^{\perp} D_k \xi^{\perp}) + \{H[\xi], \eta^i\}_{PB} - \{H[\eta], \xi^i\}_{PB} \\
	\end{split}	
\end{equation}

On a manifold with boundary there exist additional complications. Namely, the functional derivatives of the generators (A.5) are not typically well-defined \cite{ReggeAddBterm}. The simplest solution is to add a boundary term $B[\xi]$ such that the variation of $B[\xi]$ cancels the boundary variation of the Hamiltonian \cite{HamBterms}:

\begin{equation}
	\begin{split}
	\delta H[\xi] = ... - \frac{1}{16\pi G} \int_{\partial \Sigma}d^{n-2}x \Big\{ \sqrt{\sigma} &\Big[\xi^{\perp}(n^k \sigma^{\ell m} - n^m \sigma^{\ell k})D_m \delta q_{\ell k} \\
	&- D_m \xi^{\perp}(n^k \sigma^{\ell m} - n^m \sigma^{\ell k})q_{\ell k} \Big] + 2\hat{\xi}^i\delta \pi^n_{i} - \hat{\xi}^n \pi^{ij}\delta q_{ij}\Big\} 
	\end{split}	
\end{equation}
where $\sigma_{ij}$ is the induced metric on the boundary, $\sigma$ is its determinant and $n^k$ is the unit normal. However, the boundary term can alter the algebra of the generators by introducing a nonzero central extension. In \cite{EffectiveConform}, Carlip derives a general form for any possible central terms. He begins by noting that the new generators  $\widetilde{H}[\xi] = H[\xi] + B[\xi]$ have well-defined Poisson brackets by definition. One can therefore write

\begin{equation}
	\{ \widetilde{H} [\xi], \widetilde{H}[\eta]\}_{PB} = \widetilde{H}[\{\xi,\eta\}_{SD}] + K[\xi,\eta]
\end{equation}
where $K[\xi,\eta]$ represents a possible central term. Solving for $K[\xi, \eta]$ and evaluating the Poisson brackets gives

\begin{equation}
\begin{split}
	K[\xi,\eta] = -B[\{\xi,\eta\}_{SD}] &- \frac{1}{8 \pi G} \int_{\partial \Sigma}{d^{n - 2}x \sqrt{\sigma} n^k \Big[ \frac{1}{\sqrt{q}}\pi_{ik} \{\xi,\eta\}^i_{SD} - \frac{1}{2\sqrt{q}} (\hat{\xi}_k \eta^{\perp} - \hat{\eta}_k \xi^{\perp}) \mathcal{H} } \\
	&+ (D_i \hat{\xi}_k D^i \eta^{\perp} - D_i \hat{\eta}_k D^i \xi^{\perp}) - (D_i \hat{\xi}^i D_k \eta^{\perp} - D_i \hat{\eta}^i D_k \xi^{\perp}) \\
	&+ \frac{1}{\sqrt{q}}(\hat{\eta}_k \pi^{mn} D_m \hat{\xi}_n - \hat{\xi}_k \pi^{mn} D_m \hat{\eta}_n) + (\xi^{\perp} \hat{\eta}^i - \eta^{\perp}\hat{\xi}^i)^{(n-1)}R_{ik}\Big]
\end{split}	
\end{equation}	
The presence of $B[\{\xi,\eta\}_{SD}]$ on the right hand side indicates that the central term is sensitive to the choice of boundary conditions. However, as discussed by Carlip in \cite{EffectiveConform}, for the central charge of a Virasoro algebra this complication is avoidable. Suppose one obtains a one-parameter subalgebra of surface deformations

\begin{equation}
\{\xi, \eta\}_{SD} = \xi \eta' - \eta \xi'
\end{equation}
Then the boundary term depends only on this combination. On the other hand, for a Virasoro algebra, a genuine central term has a characteristic three-derivative structure $\xi'\eta'' - \eta'\xi''$, which cannot be constructed from (A.13) and its derivatives. In this case the boundary term should not contribute to the central charge.


\end{document}